\documentclass[12pt]{article} 

\usepackage{amsfonts}
 \usepackage{amssymb}
 \usepackage{amsmath}

\def\pmx{\begin{pmatrix}}
\def\emx{\end{pmatrix}}
\def\bsq{\begin{subequations}}
\def\esq{\end{subequations}}
\def\be{\begin{eqnarray}}
\def\ee{\end{eqnarray}}
\def\bee{\begin{eqnarray*}}
\def\eee{\end{eqnarray*}}
\def\bal{\begin{align}}
\def\eal{\end{align}}

\def\ds{\displaystyle}
\def\id{{\cal I}}

\newtheorem{thm}{Theorem}
\newtheorem{lemma}[thm]{Lemma}

\def\bra{\langle}
\def\ket{\rangle}
\def\dg{\dagger}
\def\kb{ \ket \bra }

\def\raw{\rightarrow}

\def\half{{\textstyle \frac{1}{2}}}

 \def\tr{\hbox{Tr} \,}
 \def\trp{\hbox{Tr} }

\def\pp{ \! +\! }
\def\ts{\textstyle}
\def\nn{\nonumber}
\def\ot{\otimes}

\def\wh{\widehat}
\def\ovb{\overline}
\def\td{\tfrac{1}{d}}
\def\qc{{\rm QC}}

\def\av{{\rm av}}

\def\pp{ \! + \! }

\def\hil{{\mathcal H}}
\def\calm{{\mathcal M}}

\def\qed{\qquad{\bf QED}}
\def\pf{ \noindent{\bf Proof:} }

\newcommand{\proj}[1]{ | #1 \kb  #1|}

\title{\large \bf Another Short  and Elementary Proof of  \\ Strong Subadditivity of
Quantum Entropy}

    \author{Mary Beth Ruskai \thanks{Partially supported  by
 the National Security Agency (NSA) and
 Advanced Research and Development Activity (ARDA) under
Army Research Office (ARO) contract number 
     DAAD19-02-1-0065, and by the National Science
        Foundation under Grant  DMS-0314228.}
      \\ {\small Department of Mathematics, 
Tufts University, 
     Medford, MA 02155 USA} \\ 
    {\small  Marybeth.Ruskai@tufts.edu}}

\begin{document}

\maketitle

  \begin{abstract}
  A short and elementary proof of the joint convexity of relative
  entropy is presented,  using nothing beyond linear algebra.    The key 
  ingredients are an easily
  verified integral representation and the strategy used to prove the
  Cauchy-Schwarz inequality in elementary courses.     Several
  consequences are proved in a way which allow an elementary
   proof of strong subadditivity in a few more lines.    
 Some expository material on 
Schwarz inequalities for operators and 
 the Holevo bound for partial measurements is also included.
 
      \end{abstract}
  
  \section{Introduction}
 
 Because the strong subadditivity (SSA) of quantum entropy plays an important
 role in quantum information theory, there has been some interest in simple
 proofs \cite{NP,RuskWYD}, suitable for elementary courses.
     In this note, we give a self-contained proof 
 of SSA, valid for finite dimensional systems, using only basic linear algebra
 and an easily verified integral representation.     The basic strategy was 
 used in \cite{LesR}.   However, the presentation here, unlike that in
 \cite{LesR} and \cite{NP}, does not explicitly use the relative modular operator.
 Instead, the simple left and right multiplication operations explained in
 Section~\ref{sect:prelim} suffice.   Unlike \cite{RuskJMP,RuskWYD} not even
 elementary results from complex analysis are used.

 The state of a quantum system is described by a density matrix $\rho$,
 i.e., a positive semi-definite matrix satisfying $\tr \rho = 1$,  The  entropy 
 of a quantum state represented by density matrix $\rho$  was defined  in 1927 by von
 Neumann \cite{vN27,vNbk}  as
 \be  \label{vNent}
     S(\rho) = - \tr \rho \log \rho.
 \ee
 The property of SSA arises when the relevant quantum system is composed
 of subsystems so that  $\rho_{ABC}$ is a density matrix on a tensor product
 space  of the form $\hil_A \ot \hil_B \ot \hil_C$, and the partial trace is used
 to define the reduced density matrices, $\rho_{AB} = \trp_C \, \rho_{ABC} $ and 
$ \rho_B = \trp_A \, \rho_{AB} = \trp_{AC} \, \rho_{ABC}$, etc.   The SSA inequality 
 \cite{SSA} is
 \be   \label{ssa}
     S(\rho_{ABC}) + S(\rho_B)  \leq   S(\rho_{AB})  + S(\rho_{BC}) .
 \ee
Many applications of SSA use closely related properties of the
relative entropy
\be  \label{relent}
    H(P,Q) = - \tr P \big( \log P - \log Q \big)
\ee
which is well-defined for positive sem-definite $P,Q$
whenever $\ker(Q) \subset  \ker(P)$ provided 
that we define  $P  ( \log P - \log Q) = 0$ on  $\ker(P)$.   
A description of the properties of $S(\rho)$ and $H(P,Q)$,
and the connections between them is given
in \cite{RuskJMP,W}.   

 The key result is the next theorem.
\begin{thm}  \label{thm:JC} 
The relative entropy is jointly convex in $P,Q$, i.e., 
when $P_j, Q_j$ are sequences of positive semi-definite matrices
satisfying $\ker(Q_j) \subset  \ker(P_j)$, then
\be   \label{eq:JC}
    H\big( \ts{ \sum_j} x_j P_j ,  \ts{ \sum_j} x_j Q_j \big) \leq  \sum_j  x_j H(P_j, Q_j)
\ee
with $x_j \geq 0$ and $\sum_j x_j = 1$.
\end{thm}
After proving Theorem~\ref{thm:JC}  in Section 2, we obtain some important
corollaries in Section~\ref{sect:JCcor}, and   
show in \eqref{ssapf} that SSA follows easily from Theorem~\ref{cor:mon}b,
without need for any auxiliary spaces or other results.     

Although our main purpose is to present a simple proof of SSA, we added some
 expository material.
  In Section~\ref{sect:CS}, we compare the argument in Section~\ref{sect:JCAQA} to
elementary proofs of the Cauchy-Schwarz inequality and 
give a direct proof of the monotonicity of relative entropy.   
In Section~\ref{sect:holv}, we present three short proofs of the Holevo bound,  
each of which is valid for partial measurements.

We will frequently use expressions, such as, $A \log Q$ or $A^{\dg} \frac{1}{Q} A$,
without requiring  the operator $Q$ to be non-singular.   But we only
do so when $\ker(Q ) \subset  \ker(A)$ and the expression
involved can be well-defined  by replacing $Q$ by $Q + \epsilon I$ and taking
a limit   $\epsilon \raw 0+$.   For simplicity and ease of exposition, we proceed as
if $Q$ is non-singular and refer to \cite{LbR2} for technical details.

  \section{Proof of joint convexity of $H(P,Q)$}
   
\subsection{Right and left multiplication}   \label{sect:prelim}

The proof  will use
 the operations of left and right multiplication by $P$ which are defined as
$L_P(X) = PX$ and $R_P(X) = XP$.  Both $L_p$ and $R_P$ are linear 
operators on the vector space of $d \times d$ matrices which becomes a Hilbert
space when equipped with the  Hilbert-Schmidt (HS)
 inner product $\bra A, B \ket = \tr A^{\dg} B $.
 The following properties are easy to verify
\begin{enumerate}

 \renewcommand{\labelenumi}{\theenumi}
    \renewcommand{\theenumi}{\alph{enumi})}
    
    \item The operators $L_P$ and $R_Q$ commute since
    \be
      L_P [R_Q(A) ] =  PAQ = R_Q[L_P(A)]
    \ee
even when $P$ and $Q$ do not commute.

\item  $L_P$ and $R_P$ are invertible if and only if $P$ is non-singular,
in which case 
 $L_P^{-1} = L_{P^{-1}}$ and $R_P^{-1} = R_{P^{-1}}$.
 
 \item  Let $\wh{L_P}$ denote the adjoint with respect to the
   HS inner product.  It follows from 
   \be   \label{HSsa}
\tr A^{\dg} L_P(B) = \tr A^{\dg} P B = \tr   (P^\dag A)^{\dg} B = \tr  [L_{P^\dag}(A)]^{\dg} B.
\ee
that  $\wh{L_P} = L_{P^\dag}$ and, similarly, $\wh{R_P} = R_{P^\dag}$.
Thus,  $P = P^{\dg}$ implies that the operators $L_P$ and $R_P$ are self-adjoint 
 
 \item  When $P \geq 0$, the operators $L_P$ and $R_P$ are positive 
 semi-definite, i.e., 
  \bee
    \tr A^{\dg} L_P(A) =   \tr A^{\dg} P(A) \geq 0 \qquad \text{and} \\
          \tr A^{\dg} R_P(A) =   \tr A^{\dg} A P = \tr  A P A^{\dg} \geq 0.
 \eee 

\end{enumerate}

% Remark about functions of operators ...

\subsection{Strategy}

We reduce the proof of the joint convexity of $H(P,Q)$ to the proof of the
following two statements.
    \begin{itemize}
  
  \item[I) ] One can write the relative entropy in the form
  \be  \label{RE.intrep}
    H(P,Q) = \int_0^{\infty} \tr (Q-P)  \frac{ 1}{L_Q   + t R_P} (Q - P) \frac{1}{(1 + t)^2} dt   
  \ee
   
  \item[II)] The map 
  $(A,P,Q) \mapsto \tr A^{\dg}  \frac{ 1}{L_Q   + t R_P} A$ is jointly
  convex in $A,P,Q$.    
     \end{itemize}
Letting $A = P-Q$ and using (II) in (I), yields
the joint convexity of $H(P,Q)$.     

Note that $\bra \phi , Q_j \, \phi \ket > 0$ for each $j$ implies
      $\bra \phi , \sum_j Q_j \, \phi \ket > 0$ so that
     $\ker(Q_j) \subset  \ker(P_j)$, 
          implies $\ker(\sum_j Q_j) \subset  \ker(\sum_j P_j)$.
   Thus,  under the hypothesis
      of Theorem~1,   all   
          expressions which arise are  well-defined.

\subsection{Proof of the integral representation I.}    \label{sect:int}

We begin with the easily verified integral representation
  \bal  \label{intrep1}
  - \log w & =   \int_0^{\infty} \bigg[ \frac{1}{w + t} -  \frac{1}{1 + t}\bigg] dt  \\
  \intertext{which can be rewritten as}  \label{intrep2}
    & =   (1-w) + \int_0^{\infty}\frac{(w-1)^2}{w + t} \frac{1}{(1 + t)^2} dt
\end{align} 
Next, use   a basis in which $Q$ is diagonal, to see that
\be
\tr ( \log L_Q)(P) = \tr   L_{ \log Q}(P) = \tr P \log Q.
\ee
  Using this and the
 fact that $L_Q$ and $R_P$ commute, one finds
  \be   \label{start1}
    H(P,Q) & = & - \tr (\log R_P^{-1})(P) -  \tr (\log L_Q)(P)   \\  
        & = & - \tr [\log \big(L_Q R_P^{-1} \big)\big](P)  \label{start2}  \nn \\
        & = &  \tr (1 - L_Q R_P^{-1})(P) +    \label{start3}    \\   \nn 
              & & ~+    \int_0^{\infty} \tr (L_Q R_P^{-1}-1)
            \frac{ 1}{L_Q R_P^{-1} + tI} (L_Q R_P^{-1}-1)(P) \frac{1}{(1 + t)^2} dt   
    \ee
    where the last step replaced $w$ by  $L_Q R_P^{-1}$ in \eqref{intrep2}.
To see why leads to \eqref{RE.intrep}, first note that
\be
  (L_Q R_P^{-1}-1)(P) = L_Q(I) - P = Q - P.
    \ee
This can be used on the far right in  \eqref{start3} 
and also gives  $\tr (1 - L_Q R_P^{-1})(P) = \tr P - Q = 0$.
Next, use property (b) above to see that
 \be
   \tr A (L_Q R_P^{-1}-1) (B) & = & \tr A (L_Q - R_P) \circ R_P^{-1}(B)    \\ \nn 
 %    & = &  \tr  [\wh{L_Q }-\wh{R_P}) (A) ] R_P^{-1}(B)  \\ \nn
      & = &    \tr  [(L_Q - R_P)(A)]  R_P^{-1}(B)    
 \ee
Using this with $A = I$ and $B = \big(L_Q R_P^{-1} + tI\big)^{-1}(X) $ gives
\be
  \tr  (L_Q R_P^{-1}-1) (X) & = &  (Q - P) R_P^{-1}\big( L_Q R_P^{-1} + tI \big)^{-1} (X) \nn  \\
    & = & (Q - P) \frac{ 1}{L_Q   + t R_P}(X)  \label{start4}
\ee
where we used  
  $R_P^{-1}  \big(L_Q R_P^{-1} + tI\big)^{-1} = 
\big[ (L_Q R_P^{-1} + tI ) R_P \big]^{-1} =(L_Q + t R_P)^{-1}$. 
Letting $X = Q-P$ and inserting \eqref{start4}  in \eqref{start3}  yields  \eqref{RE.intrep}.   \qed

\subsection{ Proof of the joint convexity II:}  \label{sect:JCAQA}

First observe that the properties of $L_P$ and $R_q$ given   in Section~\ref{sect:prelim}
and the Hilbert-Schmidt inner product \eqref{HSsa}, facilitate the evaluation  of
such expressions as
\bee
 \lefteqn{  \tr  [(L_p \pp R_Q)^{-1/2}(A)]^\dag (L_p \pp R_Q)^{-1/2}(B)  \, = \, 
       \bra (L_p \pp R_Q)^{-1/2}(A), (L_p \pp R_Q)^{-1/2}(B) \ket  }   \hskip3.5cm 
         \\
& = &       \bra A,  (L_p \pp R_Q)^{-1}(B) \ket 
~  =   ~ \tr A^\dag  (L_p \pp R_Q)^{-1}(B)  .\qquad  \qquad 
\eee       
Now let et  $M_j = (L_{P_j} + t R_{Q_j})^{-1/2}(A_j) -  (L_{P_j} + t R_{Q_j})^{1/2}(\Lambda) $,
Then
\be  \label{eq:Schz1}
   0  & \leq & \sum_j \tr M_j^{\dg} M_j  ~ = ~ \sum_j \bra M_j , M_j \ket  \nonumber  \\   
     & = &    \sum_j \tr A_j^{\dg}  (L_{P_j}  + t R_{Q_j})^{-1}(A_j)  - 
     \tr   \big( \ts{ \sum_j } A_j^{\dg}  \big) \Lambda    
               \\    & ~  & ~ \qquad \qquad  -  \tr \Lambda^{\dg} \big( \ts{\sum_j A_j } \big)
        +    \tr \Lambda^{\dg}    \ts{ \sum_j} \big( L_{P_j}  + t R_{Q_j}) \Lambda  .   \nonumber 
         \ee
Next, observe that  for any matrix $W$,
\bee 
\sum_j  \big( L_{P_j} + t R_{Q_j})(W)  & = &   \sum_j \big( P_j W + t W Q_j \big)  
 =   \big( \ts{ \sum_j} P_j \big) W + t W   \big( \ts{ \sum_j} Q_j \big)  \\
                & = &  L_{\sum_j P_j}(W)  +  t R_{\sum_j Q_j}(W)  .
                \eee
Therefore, inserting the choice
 $\Lambda =   \big(  L_{\sum_j P_j}   +  t R_{\sum_j Q_j} \big)^{-1}  
         \big( \ts{\sum_j A_j } \big) $
in  (\ref{eq:Schz1}) yields
 \be    \label{eq:Schwzt}
  \tr   \big( \ts{ \sum_j } A_j \big)^{\dg}  \dfrac{1}{  L_{\sum_j P_j}   +  t R_{\sum_j Q_j} } 
         \big( \ts{\sum_j A_j } \big)
  \leq   \sum_j \tr A_j^{\dg}  \dfrac{1}{L_{P_j} + t R_{Q_j}}(A_j) .
            \ee
for any $t \geq 0$.     Since
$\ds{ (x A)^\dag \frac{1}{L_{xP} + R_{xQ} } (x A) = x  \Big(A^\dag \frac{1}{L_p + R_Q }(A) \Big)}$
this implies\footnote{If this is not obvious, see the Appendix.}  
 joint convexity. \qed

\subsection{Remarks}

For simplicity, we used \eqref{start1} as the starting point for obtaining
the integral representation \eqref{RE.intrep}.     It is equivalent, and customary,
to begin instead with a symmetric variant of \eqref{start2},
$H(P,Q) =   - \tr  P^{1/2} [\log \big(L_Q R_P^{-1} \big)\big](P^{1/2})  $
and then observe that  \linebreak
   $\big(L_Q R_P^{-1} - I \big)(P^{1/2}) = (R_P)^{-1/2}(Q-P)$

One advantage to our approach, like that in \cite{NP}, is that it is easily
extended to give a proof of joint convexity when $- \log w$ is replaced by
another operator convex function.   This only changes the
weight function in  the integral; see \cite{LesR,Pz2} for details.
Replacing $\frac{1}{(1+t)^2}$
by $\delta(1-t)$ in \eqref{RE.intrep} yields $(Q-P) \frac{1}{L_P + R_Q} (Q-P)$ which is the  
generalized relative entropy whose Hessian yields the Riemmanian
metric associated with the Bures metric 
$D^{\rm Bures}(P,Q)   =  \big[2 \big(1 - \tr (\sqrt{P} Q \sqrt{P})^{1/2} \big) \big]^{1/2} $.

\section{Consequences of joint convexity} \label{sect:JCcor}

\subsection{Monotonicity of relative entropy}

The joint convexity of relative entropy implies the well-known fact \cite{Lind75,OP, Pz1,Uhl2}
that it decreases  under  completely positive, trace-preserving (CPT) maps.
These maps represent quantum channels.   We will prove this
result by first considering two special cases, the partial trace 
and the projection onto the
diagonal,  which are of sufficient 
importance to deserve separate statements and have extremely
elementary proofs.
\begin{thm}  \label{cor:mon}
Let $\Phi^{\qc}$ denote the map which projects a matrix onto its diagonal,
and let $\Phi$ be any CPT map. Then

a)  $ ~~ H[\Phi^\qc(\rho), \Phi^\qc(\gamma)] \leq H(\rho, \gamma)$

b)  $ ~~ H[\rho_A, \gamma_A ] \leq H(\rho_{AB}, \gamma_{AB})$

c)  $ ~~ H[\Phi(\rho), \Phi(\gamma)] \leq H(\rho, \gamma)$

\end{thm}
\pf First, let $Z$ denote the diagonal unitary matrix with elements
$z_{jk} = \delta_{jk} \omega^k$ with $\omega =  e^{i 2 \pi/d } $ and 
note that that $(1 -\omega^{(k-n)}) \sum_j \omega^{j(k-n)} = 1 - \omega^{dj(k-n)} = 0$.
Then, for any matrix $X$
\be
  \sum_j Z^j  X Z^{-j}  =    \sum_j \omega^{j(k-n)} x_{kn} =  d \, \delta_{kn} x_{kn} 
  \ee
which implies that 
$\Phi^\qc(\rho) \equiv \td   \sum_j Z^j  X Z^{-j}$ projects a matrix onto its diagonal.

Now write a bipartite state $\rho_{AB} = \sum_{jk} |e_j \kb e_k | \ot P_{jk}$ as a block matrix
with blocks $P_{jk}$.   Then
\be   \label{corpfa}
   H( \rho_B,\gamma_B)  & = &  H\big(  \ts{ \sum_k P_{kk} \, , \sum_k Q_{kk} \big)  ~ \leq  ~
      \sum_k      H( P_{kk}, Q_{kk} })\\ \nn 
         & = &  H\big( \ts{ \sum_k \proj{e_k} \ot P_{kk}\,  , \, \sum_k \proj{e_k} \ot  Q_{kk}}  \big) \\  \label{corpfb}
        &= & 
    H\big[ (\id_A \ot  \Phi^\qc)(\rho_{AB}),(\id_A \ot  \Phi^\qc)(\gamma_{AB}) \big]  \\ \nn 
       & = &  H\Big[  \td \sum_j  (I \ot Z)^j \rho_{AB} (I \ot Z)^{-j}, \, \td \sum_j  (I \ot Z)^j \gamma_{AB} (I \ot Z)^{-j}  \Big]   \\ \nn 
       & \leq & \td \sum_j    H\big[ (I \ot Z)^j \rho_{AB} (I \ot Z)^{-j},  (I \ot Z)^j \gamma_{AB} (I \ot Z)^{-j}\big] \\
       & = & H[\rho_{AB}, \gamma_{AB}]   \label{corpfc}
     \ee  
 where Theorem~1 was used twice in the subadditive form \eqref{JCsub}, 
 and the final equality uses the fact that
 conjugation of both arguments by a unitary matrix does not change $H(\rho,\gamma)$.
 This proves part (b).
 When the space $\hil_A$ is 1-dimensional,
 the inequality between \eqref{corpfb} and \eqref{corpfc} yields part (a).
 
 To prove (c) fix the ancilla representation of Lemma~\ref{ancrep} and let
\bee
\sigma_{AB} \equiv  U_{AB} \, \rho \ot  \proj{\phi_B} \, U_{AB}^{\dg}  \qquad
  \tau_{AB} \equiv U_{AB} \, \gamma \ot  \proj{\phi_B} \, U_{AB}^{\dg}.
\eee  
Then $\Phi(\rho) = \sigma_A$, $\Phi(\gamma) = \tau_A$ and, since $U_{AB}$
is unitary, $H(\rho,\gamma) = H(\sigma_{AB}, \tau_{AB})$.   Thus, 
it follows from part (b) that
\be
    H[\Phi(\rho), \Phi(\gamma)]  =  H(\sigma_A, \tau_A)  
              \leq   H(\sigma_{AB}, \tau_{AB} = H(\rho,\gamma).   \qed
\ee
  
\subsection{Convexity corollaries}
The conditional entropy is given by
\be  \label{condent}
 S(\rho_{AB}) -   S(\rho_A)   =  - H(\rho_{AB}, \rho_A \ot \tfrac{1}{d} I) + \log d,
\ee
It then follows immediately from the joint convexity of $H(\rho,\gamma)$ that
\be  \label{PTconv}
\rho_{AB} \mapsto  S(\rho_{AB}) -   S(\rho_A)  \qquad \text{ is concave}.
\ee
Moreover, for any CPT map $\Phi$, the map
\be   \label{CPTconv}
\rho  \mapsto S(\rho) - S[\Phi(\rho)]  \qquad \text{ is concave}.
\ee
This follows from \eqref{PTconv}.   Use the same 
notation as in part (c) of the previous section and observe that
$  S(\rho) - S[\Phi(\rho)]  = S(\sigma_{AB} ) - S(\sigma_B) $.

\subsection{Completing the proof of SSA}  \label{sect:SSApf}

The SSA inequality \eqref{ssa} follows immediately from Corollary~\ref{cor:mon}b
with  $\gamma =  \rho_{AC} \ot \tfrac{1}{d} I$.    We write this out explicitly using
 \eqref{condent}.
\be   \label{ssapf}
S(\rho_A)  - S(\rho_{AB})     & = &    H(\rho_{AB}, \rho_A \ot \tfrac{1}{d} I) - \log d \nn \\
    & \leq &  H(\rho_{ABC}, \rho_{AC} \ot \tfrac{1}{d} I) - \log d  \qquad \\ \nn
    & = &  S(\rho_{ABC}) - S(\rho_{AB})  .    \qquad    \qquad  \qed
\ee 

There is another form of SSA which follows easily from \eqref{PTconv}, namely,   
\be \label{ssa:alt}
    S(\rho_B)  + S(\rho_D) \leq   S(\rho_{AB}) + S(\rho_{AD}).
\ee
  To prove this 
first consider  
\be
      F(\rho_{ABD}) = S(\rho_{AB}) + S(\rho_{AD}) - S(\rho_B)  - S(\rho_D) .      
\ee
When $\rho_{ABD}$ is pure, it follows from  Lemma~\ref{pure2} that
$S(\rho_{AB}) = S(\rho_D)$ and $ S(\rho_{AD}) = S(\rho_B) $.
Thus,   $F(\rho_{ABD}) = 0$ for pure states.   Since  $F(\rho_{ABC})$ is the sum
of two  functions $S(\rho_{AB}) - S(\rho_B) $ and $+ S(\rho_{AD}) - S(\rho_D)   $
which are concave by \eqref{PTconv}, the map $\rho_{ABD} \mapsto F(\rho_{ABD}) $
is also concave.
 Since any mixed  $\rho_{ABD}$  is a convex
combination of pure states, $F(\rho_{ABD}) \geq  0$, which implies \eqref{ssa:alt}.

By  Lemma~\ref{pure1},  one can  purifiy $\rho_{ABC}$ or   $\rho_{ABD}$ to $\rho_{ABCD}$ and use
Lemma~\ref{pure2}  to show that \eqref{ssa:alt}   holds if and only if \eqref{ssa} does.

  \section{Remarks on Cauchy-Schwarz type inequalities} \label{sect:CS}
  
  \subsection{Elementary proof strategy}

The elementary vector version of the Cauchy-Schwarz inequality states that
\be   \label{CSelem}
\big| \sum_k \ovb{v}_k w_k \big|^2 \leq  \Big( \sum_k  |v_k|^2 \Big) \Big( \sum_k  |w_k|^2 \Big)
\ee
When $v_k = p_k^{1/2} ,    w_k =   p_k^{ -1/2} a_k$, this can be written as
\be  \label{CSel2}
\ts{  \sum_k}  \ovb{a}_k   \dfrac{1}{\ts{  \sum_k} p_k  }  \ts{  \sum_k}  a_k    \leq   \sum_k \ovb{a}_k \dfrac{1}{p_k} a_k.
\ee
In \cite{LbR2}, Lieb and Ruskai proved an operator version of \eqref{CSel2}, namely that
\be  \label{CSop}
  \ts{ \sum_k } A_k^{\dg} \,   \dfrac{1}{ \ts{ \sum_k } P_k } \,     \sum_k   A_k  
           \leq   \sum_k A_k^{\dg}  \dfrac{1}{P_k}  A_k.
\ee
holds as an operator inequality.    This is equivalent to the statement that the
map $(A,P) \mapsto A^{\dg} P^{-1} A$ is jointly operator convex.  
The proof in Section \ref{sect:JCAQA} is based on that in \cite{LbR2} which (although 
published later) actually preceded the proof of SSA.     However, without 
the additional ingredient of $L_P$ and $R_Q$, which are motivated
by Araki's subsequent introduction \cite{Ak} of the relative modular operator,
the results in \cite{LbR2} are not sufficient to prove SSA.    The
recognition that the argument in \cite{LbR2} could be modified to prove
SSA took another 25 years \cite{LesR}.

The proofs in both \cite{LbR2}  and Section \ref{sect:JCAQA}
are variants of the standard strategy used to prove the elementary
inequality \eqref{CSelem}.    One observes that  $ \sum_k |v_k + \lambda w_k|^2 \geq 0$
and shows that the minimizing choice  $\lambda = -  (\sum_k v_k)/ (2 \sum_k w_k)$
yields   \eqref{CSelem}.    In Section \ref{sect:JCAQA} the operator $\Lambda$
plays the role of $\lambda$.

\subsection{Schwarz  inequalities for CP maps}

We now consider Schwarz type inequalities involving completely positive
(CP) maps.  Let $\Phi$ be a CP map written in Kraus form 
$\Phi(P) = \sum_j K_j P K_j^{\dg}$.    It then follows from  \eqref{CSop} that
\be  \label{CSCP}
   [\Phi(A)]^{\dg} \frac{1}{\Phi(P)} \Phi(A) & = &   \nn
       \sum_j  K_j A^{\dg} K_j^{\dg} \frac{1}{\sum_j K_j P K_j^{\dg}}  \sum_j K_j A K_j^{\dg} \\
       & \leq & \sum_j K_j A^{\dg} \frac{1}{P} AK_j^{\dg}  
       =  \Phi\Big( A^{\dg}  \frac{1}{P} A \Big).
\ee
By making the replacements 
$A \raw B^{\dg} A$ and $P \raw B^{\dg} B$ in \eqref{CSCP}
one finds
\be   \label{CSAB}
    \Phi(A^{\dg} B) \frac{1}{\Phi(B^{\dg} B)} \Phi(B^{\dg} A) \leq \Phi(A^{\dg} A)
\ee
This inequality  is  proved  in  \cite{LbR2} using the Stinespring 
\cite{Paul,Stine} representation..

Choi \cite{Choi1} realized that    \eqref{CSCP} and \eqref{CSAB}
hold under the weaker condition that $\Phi$ is 2-positive.  His approach is quite
different, and based on the fact that a $2 \times 2$
  block matrix $ \pmx P & C \\ C^{\dg} & Q \emx$
is positive semi-definite if and only if $C = \sqrt{P} X \sqrt{Q}$ with $X$
a contraction, i.e., $X^{\dg} X \leq I$.    When $P,Q$ are both non-singular,
this is equivalent to $C^{\dg} P^{-1} C \leq Q$.     The $2$-positivity of $\Phi$ 
says that
 $\pmx \Phi(A^{\dg}A) & \Phi(A^{\dg}B) \\ \Phi(B^{\dg}A) & \Phi(B^{\dg}B) \emx$
 is positive semi-definite.
Applying the   condition above yields \eqref{CSAB}.

\subsection{Monotonicity of relative entropy}

In \cite{LesR} a   strategy similar to that in Section~\ref{sect:JCAQA}
was used to give a direct proof of the montonicity
of relative entropy under CPT maps without using an auxiliary space.    
One begins as before, but with
$M = (L_P + t R_Q)^{-1/2}(A) - (L_P + t R_Q)^{1/2}[\wh{\Phi}(X)] $ and 
$X =\big( L_{\Phi(P)} + t R_{\Phi(Q)} \big)^{-1}[ \Phi(A)]$.  Then $M^{\dg}M \geq 0$
implies
\be  \label{monopf1}
 \lefteqn{  \tr A^{\dg} \frac{1}{L_P + t R_Q} A - 
       2 \tr [\Phi(A)]^{\dg} \frac{1}{L_{\Phi(P)} + t R_{\Phi(Q)}} \Phi(A)} \hskip4cm \\
    & ~ &    + ~  \tr [\wh{\Phi}(X)]^{\dg} [L_P + t R_Q] \wh{\Phi}(X) \geq 0 \nn
\ee
Comparing the last two terms requires a bit more work and the use  of
 \eqref{CSAB}.  Since $\Phi$ trace preserving implies
 $\wh{\Phi}(I) = I$,   \eqref{CSAB} implies
 $ [\wh{\Phi}(X)]^{\dg} \wh{\Phi}(X) \leq \wh{\Phi}(X^{\dg} X)$ and
$\wh{\Phi}(X)[\wh{\Phi}(X)]^{\dg}  \leq \wh{\Phi}(X X^{\dg})$ .  Then, using
the cyclicity of the trace, one finds
\be
\tr [\wh{\Phi}(X)]^{\dg} [L_P + t R_Q] \wh{\Phi}(X) & = & 
  \tr \wh{\Phi}(X)  [\wh{\Phi}(X)]^{\dg} P + t \tr [\wh{\Phi}(X)]^{\dg}   \wh{\Phi}(X) Q
    \quad  \nn \\
  & \leq  & \tr  \big[ \wh{\Phi}(X X^{\dg}) P + t \wh{\Phi}(  X^{\dg}X) Q \big] \\
  & = &  \tr  \big[ X X^{\dg} \Phi(P) + t X^{\dg}X \Phi(Q) \big]  \nn  \\
  & = & \tr X^{\dg} \big[ L_{\Phi(P)} + t R_{\Phi(Q) } \big] X \\ \nn 
  & = &  [\Phi(A)]^{\dg} \frac{1}{L_{\Phi(P)} + t R_{\Phi(Q)}} \Phi(A)  
  \ee
Using this in \eqref{monopf1} allows one to combine the last two terms
as before.    Substituting the resulting inequality in \eqref{RE.intrep}  yields
part (c) of Theorem~2.

\section{Holevo bounds for partial measurements}  \label{sect:holv}

In order to state the Holevo bound, we introduce some notation.
Let ${\cal E}$ denote an ensemble $\{ \pi_j, \rho_j \}$ with $\pi_j > 0,
\sum_j \pi_j = 1$ and each $\rho_j$ a density
matrix.    The Holevo $\chi$-quantity is defined as
\be   \label{Holvbd}
   \chi({\cal E}) = S\Big( \sum_j \pi_j \rho_j \Big) - \sum_j \pi_j  S(\rho_j).
\ee
A set of positive semi-definite operators $\{ M_a \}$ satisfying $\sum_a M_a$
is called a positive operator valued measurement (POVM) and denoted ${\cal M}$.   
Every POVM defines a CPT map $\Phi_{\cal M}$ which takes
$\rho \mapsto \sum_a (\tr \rho M_a) \proj{a}$.     The usual Holevo bound
states that
\be   \label{hvbd}
 \chi({\cal E}) \geq  \chi\big[ \Phi_{\calm}({\cal E}) \big] \equiv 
   S\Big[ \sum_j \pi_j \Phi_{\calm}(\rho_j )\Big] - \sum_j \pi_j   \Phi_{\calm}(\rho_j ).
\ee
where $\Phi_{\calm}({\cal E})$ denotes the ensemble in which each
$\rho_j$ is replaced by  $\Phi_{\calm}(\rho_j)$.

There is now an extensive literature  on bounds involving partial measurements.
Consider  the situation in which two parties, Alice and
Bob, share an ensemble of (possibly entangled) states $\{ \pi_j, \rho^{AB}_j \}$
on $\hil_A \ot \hil_B$, on which  one of the parties makes a measurement.    
In such cases, one expects a bound of the form
\be   \label{HolvAB}
    \chi({\cal E}^{AB}) \geq  \chi\big[ (I \ot \Phi_{\calm_B})({\cal E^{AB}}) \big]   \geq 
       \chi\big[ ( \Phi_{\calm_A} \ot \Phi_{\calm_B})({\cal E^{AB}}) \big] .
\ee
We observe that three simple strategies for proving \eqref{hvbd}
 easily extend to \eqref{HolvAB}.

The first  proof uses the  observation of Yuen and Ozawa \cite{YO} that
\be
S(\rho_\av) - \sum_j \pi_j S(\rho_j) = \sum_j \pi_j H(\rho_j, \rho_\av).
\ee
where $\rho_\av = \sum_j \pi_j \rho_j$.  
It follows from part (c) of Theorem~\ref{cor:mon} that
\be
 \lefteqn{  H[(\Phi_{\calm_A} \ot \Phi_{\calm_B})(\rho^{AB}_j), 
     (\Phi_{\calm_A} \ot \Phi_{\calm_B})(\rho^{AB}_\av)] } \qquad     \\
     & \leq & 
      H[I \ot \Phi_{\calm_B})(\rho^{AB}_j), (I \ot \Phi_{\calm_B})(\rho^{AB}_\av)]  
      ~ \leq ~ H(\rho^{AB}_j, \rho^{AB}_\av)  \nn
\ee
which is equivalent to \eqref{HolvAB}.

The next proof uses the fact that  $\chi({\cal E})$ can be regarded as a form
of mutual information between the quantum states $\rho_j$ and their
classical probability distribution $\pi_j$.   Let
$\gamma_{QC} = \sum_j \pi_j \rho_j \ot \proj{j}$  be a density matrix
on $\hil_Q \ot \hil_C$.  Then, as was observed in \cite{KR2},
\be
\chi({\cal E})  = S(\rho_\av) - \sum_j \pi_j S(\rho_j) = 
    H\big(\gamma_{QC}, \gamma_Q \ot \gamma_C \big).
\ee
Then  part (c) of Theorem~\ref{cor:mon} gives
\be
H\big[ (\Phi \ot I) (\gamma_{QC}),  (\Phi \ot I) (\gamma_Q \ot \gamma_C) \big]
  \leq H( \gamma_{QC}, \gamma_Q \ot \gamma_C).
\ee
which is equivalent to \eqref{hvbd}. 
To obtain  \eqref{HolvAB}, let ${\cal H}_Q = {\cal H}_A \ot {\cal H}_B$ and observe that
\be
\lefteqn{H\big[ (\Phi \ot \Phi \ot I) (\gamma_{ABC}),  (\Phi \ot   \Phi \ot I) (\gamma_{AB} \ot \gamma_C) \big]}    \qquad  \nn \\
     & \leq &
     H\big[ (\Phi \ot I \ot I) (\gamma_{ABC}),  (\Phi \ot  I \ot I) (\gamma_{AB} \ot \gamma_C) \big]  \\ \nn 
      & \leq & H( \gamma_{ABC}, \gamma_{AB} \ot \gamma_C).
\ee

The final proof uses the observation in  \cite{LbS},
 that the Holevo bound \eqref{Holvbd} is equivalent to the 
statement that
 $\rho \mapsto S(\rho) - S[\Phi_{\calm}(\rho)]$ is convex, which is a special case of
 \eqref{CPTconv}.   Thus,  
 the bound \eqref{HolvAB} follows immediately  from  \eqref{CPTconv}
 with $\Phi$ replaced first by  $I_A \ot \Phi_{\calm_B}$ and 
 then by $\Phi_{\calm_A} \ot \Phi_{\calm_B}$.

%  \pagebreak

 \appendix  \section{Appendix}
 
 Let $A = \sum_k \lambda_k \proj{\phi_k}$ be a self-adjoint matrix with eigenvalues
 $\lambda_k$ in the domain of the function $f(w)$.   Then we define
 $f(A) =  \sum_k  f(\lambda_k) \proj{\phi_k}$.    This is equivalent to any
 other reasonable definition and  implies that
 substituting $L_Q R_P^{-1}$ for $w$ to obtain \eqref{start3} 
 is fully justified;  there is no need to explicitly find 
 the eigenvalues and eigenvectors of $L_Q R_P^{-1}$.
 
If a function $F$  satisfies $F(x A) = x F(A)$ then convexity is equivalent to
subadditivity.   First, observe that when $F$ is also
convex
 \be
     \half F(A+B) =  g \big(\half[A+B] \big) \leq \half F(A) + \half F(B) .
 \ee
 Conversely, if $F$ is   subadditive, then
 \be
     F\big[x A + (1-x) B\big] \leq  F(xA) + F[(1-x)B] = x F(A) + (1-x) F(B).
 \ee
 Although the relative entropy $H(P,Q)$ is usually considered for
 density matrices, \eqref{relent} defines it more broadly.
Since Klein's inequality \cite{NC,OP,RuskJMP} says that
 $H(P,Q) \geq \tr P - \tr Q$, it follows that $H(P,Q) \geq 0$ when
 $\tr P = \tr Q$.  It is easy to verify that  $H(xP,xQ) = x H(P,Q)$ for
 $x > 0$.   Therefore, by the observations above, \eqref{eq:JC} is
 equivalent to
 \be  \label{JCsub}
    H\big( \ts{ \sum_j}  P_j ,  \ts{ \sum_j}  Q_j \big) \leq  \sum_j   H(_j P_j, Q_j).
 \ee

For completeness, we also state some well-known results used in 
proving corollaries to the joint convexity and SSA.
None are needed to obtain a proof of SSA.
 
\begin{lemma}  \label{ancrep}  {\em (Ancilla representation)}
Any CPT map $\Phi:  M_d \mapsto M_d$ can be 
represented  using an auxiliary space $\hil_B$ as
 \be
 \Phi(\rho) = \trp_B  U_{AB} \, \rho \ot  \proj{\phi_B} \, U_{AB}^{\dg}
 \ee
where $U_{AB}$ is unitary and $ \proj{\phi_B}$ is a pure state.
If $\sigma_{AB} \equiv U_{AB} \, \rho \ot  \proj{\phi_B} \, U_{AB}^{\dg}$,
then  $\Phi(\rho) = \gamma_A$ and $S(\sigma_{AB}) = S(\rho)$.
\end{lemma}
This is essentially a corollary to the Stinespring
representation theorem \cite{Stine}.   It was introduced in the form mused here by
Lindblad \cite{Lind75} who made the observation  about entropy and used
it to give the first proof of Theorem~\ref{cor:mon}c.
For an  overview of representation
theorems, see Chapter~2  of  Paulsen \cite{Paul}; for short accessible summaries,
see the appendices to  \cite{HvCC,KMNR,KW} as well
as Section~III.D   of \cite{RuskJMP}.  The term ``ancilla 
representation'' is introduced in \cite{KW}.

The following well-known, and
easily proved, facts go back at least to \cite{AL}.   For
further references and discussion see \cite{KR1,NC,RuskJMP}.
\begin{lemma} \label{pure2}
 When $\rho_{AB} = \proj{\psi_{AB}}$ is a pure state,  its reduced density
 matrices $\rho_A$ and $\rho_B$ have the
 same non-zero eigenvalues and  $S(\rho_A) = S(\rho_B)$. 
 \end{lemma}
 
\begin{lemma} \label{pure1}
Given a density matrix $\rho$ in $M_d$ of rank $m$, one can find
 a  pure $\rho_{AB}$ in $M_d \ot M_m$ with $\rho_A = \rho$.
\end{lemma}

  \end{document}